\documentclass[twocolumn,preprintnumbers,amsmath,amssymb,showpacs,aps,groupedaddress]{revtex4}
\usepackage{graphicx}
\usepackage{dcolumn}
\usepackage{bm}
\usepackage{amsmath}
\usepackage{amssymb}
\usepackage{epsf}
\usepackage{epstopdf}
\usepackage{multirow}
\usepackage{color}
\definecolor{Red}{rgb}{1.0,0,0}
\definecolor{Blue}{rgb}{0,0,1.0}
\newcommand{\be}{\begin{equation}}
\newcommand{\ee}{\end{equation}}
\newcommand{\bea}{\begin{eqnarray}}
\newcommand{\eea}{\end{eqnarray}}

\begin{document}
\title{Reversing Hydride Ion Formation in Quantum Information Experiments with Be$^+$}
\author{Brian C. Sawyer}
\email{brian.sawyer@boulder.nist.gov}
\author{Justin G. Bohnet}
\author{Joseph W. Britton}
\author{John J. Bollinger}
\affiliation{Time and Frequency Division, National Institute of Standards and Technology, Boulder, CO  80305}

\begin{abstract}
We demonstrate photodissociation of BeH$^+$ ions within a Coulomb crystal of thousands of $^9$Be$^+$ ions confined in a Penning trap. Because BeH$^+$ ions are created via exothermic reactions between trapped, laser-cooled Be$^+$($^2\text{P}_{3/2}$) and background H$_2$ within the vacuum chamber, they represent a major contaminant species responsible for infidelities in large-scale trapped-ion quantum information experiments. The rotational-state-insensitive dissociation scheme described here makes use of 157 nm photons to produce Be$^+$ and H as products, thereby restoring Be$^+$ ions without the need for reloading. This technique facilitates longer experiment runtimes at a given background H$_2$ pressure, and may be adapted for removal of MgH$^+$ and AlH$^+$ impurities.
\end{abstract}


\maketitle

Crystals of laser-cooled atomic ions form the basis of many experimental realizations of quantum logic gates~\cite{Ospelkaus11,Monz11,Lin13} and simulations of quantum many-body physics~\cite{Richerme14,Jurcevic14,Britton12}. In room-temperature vacuum systems, collisions with neutral background gas molecules limit quantum coherences and quantum logic operations~\cite{Bible}. Here we focus on inelastic (i.e. reactive) collisions where exothermic chemical reactions with background gas molecules can generate, in both Penning and rf traps, co-trapped molecular ions over a wide range of trapping conditions and crystal geometries~\cite{Bible,Roth06,Bertelsen04,Britton12}. These contaminants alter ion-crystal eigenfrequencies and can complicate quantum logic interactions~\cite{McAneny13}. Impurities may be resonantly ejected from both radiofrequency and Penning traps, but this technique is least efficient when the impurities are similar in mass to the qubit ion~\cite{Grosshans90}. The rate of impurity ion formation increases linearly with the number of ion qubits in the register, and the time required for loading new ions and recalibrating the register increases with the size of the system. Therefore, experimental techniques to prevent or reverse qubit-ion chemistry will be a key tool for many-ion quantum information platforms. In this article, we experimentally demonstrate a rotationally-insensitive photodissociation (PD) technique for BeH$^+$ accumulated within a $^9$Be$^+$ Coulomb crystal. Beryllium ions have found frequent use in quantum information experiments because their light mass results in tight trap confinement.

Photodissociation of molecular ions is a topic of significant interest in the atomic and chemical physics communities. Recent experimental demonstrations with trapped heteronuclear diatomic molecular ions include MgH$^+$~\cite{Bertelsen04,Kahra12}, CaH$^+$/CaD$^+$~\cite{Hansen14}, HD$^+$~\cite{Shen12}, AlH$^+$~\cite{Seck14}, HfF$^+$~\cite{Ni14}, SrCl$^+$~\cite{Puri14}, and BaCl$^+$~\cite{Chen11}. Given the difficulty of detecting scattered photons from dilute molecular-ion samples, rotational-state-selective PD has emerged as a critical tool for molecular ion spectroscopy~\cite{Seck14,Versolato13} and precision measurements~\cite{Bressel12,Ni14}, while rotationally-insensitive dissociation schemes provide valuable tests of molecular theory~\cite{Chen11,Puri14,Hansen14}. The work described here is the first experimental photodissociation of BeH$^+$.

\begin{figure}[t]
\resizebox{8.5cm}{!}{
\includegraphics{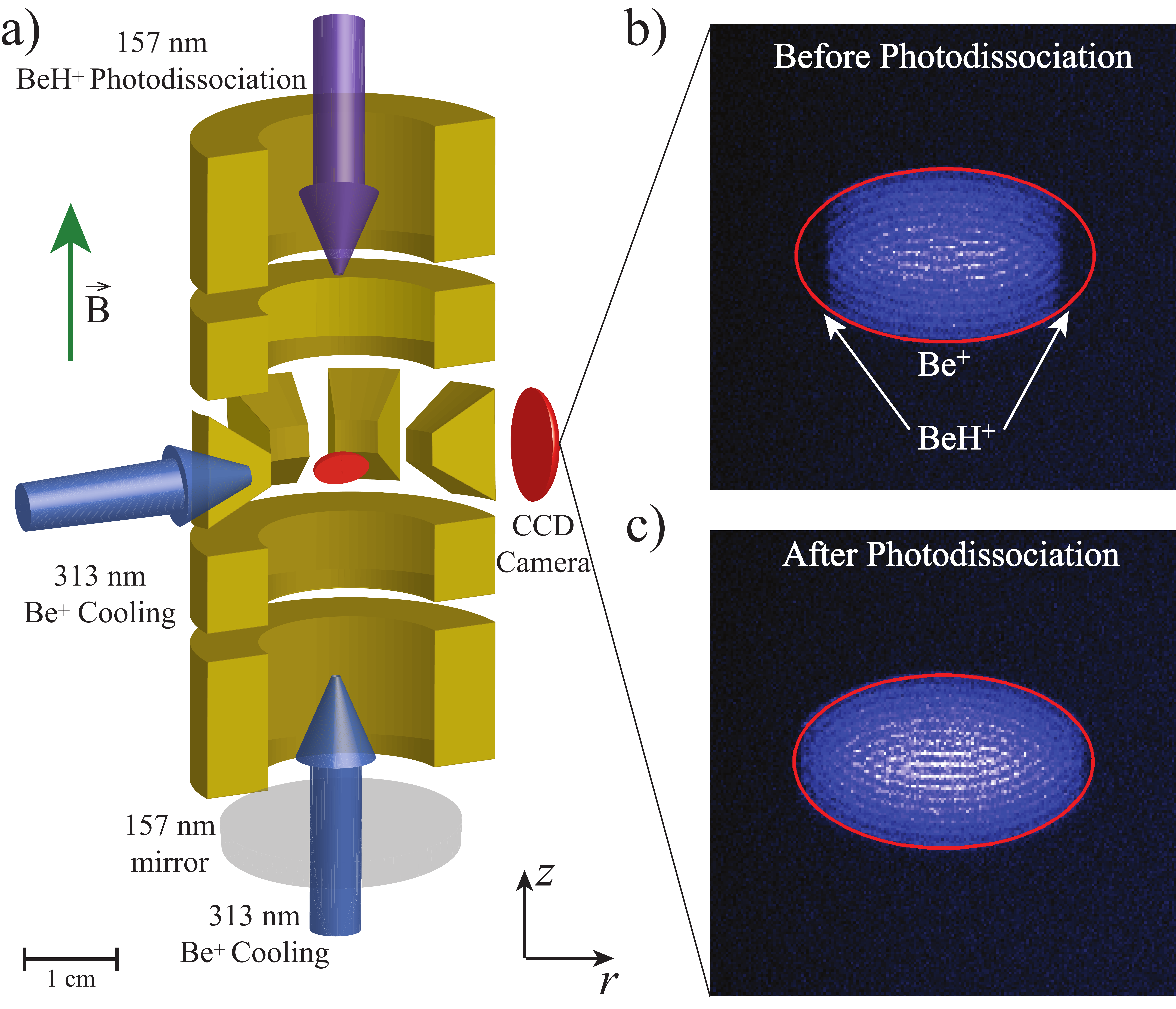}}
\caption{\label{trap}(color online) (a) Illustration of the experiment trap electrode assembly in cross section with routing of laser beams for Doppler cooling of Be$^+$ and photodissociation of BeH$^+$. A spheroidal ion cloud is depicted at the center of the trap. The vertical axis of the electrode stack is aligned to the external, uniform magnetic field of 4.46 T. (b,c) Side view images of the trapped Coulomb crystal of $\sim$1.2 $\times 10^4$ ions (a) before and (b) after 9000 pulses of 157 nm photodissociation light at 1.6 mJ/pulse. We measure an increase in the Be$^+$ fraction from $81\%$ to $97\%$ of the constant total ion number in the crystal. Each image is 1.2 mm $\times$ 1.2 mm.}
\end{figure}

Figure~\ref{trap}(a) shows an illustration of the Penning trap used for this work. The trap assembly is installed in the room-temperature bore of a superconducting magnet that produces a uniform, vertical magnetic field ($B$) of $\sim$4.46 T. This magnified view of the experiment trap region consists of 12 Au-coated Ti electrodes; eight are visible in cross section. Ions are confined in the vertical ($z$) direction by a quadrupole electric ($E$) field generated through application of static voltages to the trap electrodes. Radial ($r$) confinement results from $\vec{E} \times \vec{B}$-induced rotation through the magnetic field. In a frame of reference rotating with the ions, confinement is described by
\begin{eqnarray}
q\Phi_{\text{rot}}(r,z) &=& \frac{1}{2}M\omega_z^2 \left(z^2 + \beta_r r^2 \right), \label{trappot} \\
\beta_r &\equiv& \frac{\omega_r(\Omega_c - \omega_r)}{\omega_z^2} - \frac{1}{2}, \label{beta}
\end{eqnarray}
where $\omega_z$ is the axial confinement frequency, $q$ ($M$) is the ion charge (mass), $\Omega_c$ is the cyclotron frequency, and $\omega_r$ is the ion rotation frequency. For this work, we obtain an axial frequency of $\omega_z \sim 2\pi \times 900$ kHz by applying a combination of 0 V to the top and bottom rings of Fig.~\ref{trap}(a), -400 V to all central segmented electrodes, and -200 V to the two interior ring electrodes. The ion-electrode distance is 1 cm. We apply a radial dipole `rotating wall' potential to the azimuthally-segmented central electrodes to control the rotation frequency of the ion cloud, thereby controlling the aspect ratio of the trap potential according to Eqs.~(\ref{trappot}) and~(\ref{beta})~\cite{Huang98}. With $\omega_z=2\pi\times 900$ kHz and $\Omega_c = 2\pi \times 7.6$ MHz, $^9$Be$^+$ ions experience radial confinement for $53.7\text{ kHz}< \omega_r/2\pi < 7.55$ MHz. For all experimental data presented here, we operate with $\omega_r = 2\pi \times110$ kHz.

Our Doppler laser cooling scheme for $^9$Be$^+$ is as described in~\cite{Britton12,Sawyer12,Sawyer14}. Briefly, we apply $\sim$313-nm laser beams along both the axial and radial trap dimensions to produce a Coulomb crystal. The cycling transition between the $|J=1/2,m_J=+1/2\rangle$ ($^2$S) and $|3/2,+3/2\rangle$ ($^2$P) states yields a Doppler cooling limit of $\sim$1 mK, where $J$ and $m_J$ are the total electronic angular momentum and its projection along the magnetic field axis, respectively. This laser cooling leads to the formation of BeH$^+$ impurities. Since the reaction Be$^+$($^2$S)+H$_2$ $\rightarrow$ BeH$^+$ + H is endothermic by 1.57 eV, only Be$^+$ in an excited (e.g. $^2$P) state can react to form BeH$^+$~\cite{Raimondi83}. As a result, the reaction rate may be varied by adjusting the saturation parameter, $s = I_c/I_{sat}$, of the Be$^+$ cooling laser~\cite{Roth06}. Here $I_c$ is the cooling laser intensity and $I_{sat} \sim 300$ mW cm$^{-2}$ is the saturation intensity of the cycling transition assuming linear polarization perpendicular to the trap $B$-field. Previous work has shown that the reaction rate ($k_L$) between Be$^+$($^2$P) and H$_2$ is in good agreement with estimates from a Langevin capture model~\cite{Roth06}, which gives $k_L \sim 1.6 \times 10^{-9} \text{ cm$^3$ s$^{-1}$}$.

\begin{figure}[t]
\resizebox{7.5cm}{!}{
\includegraphics{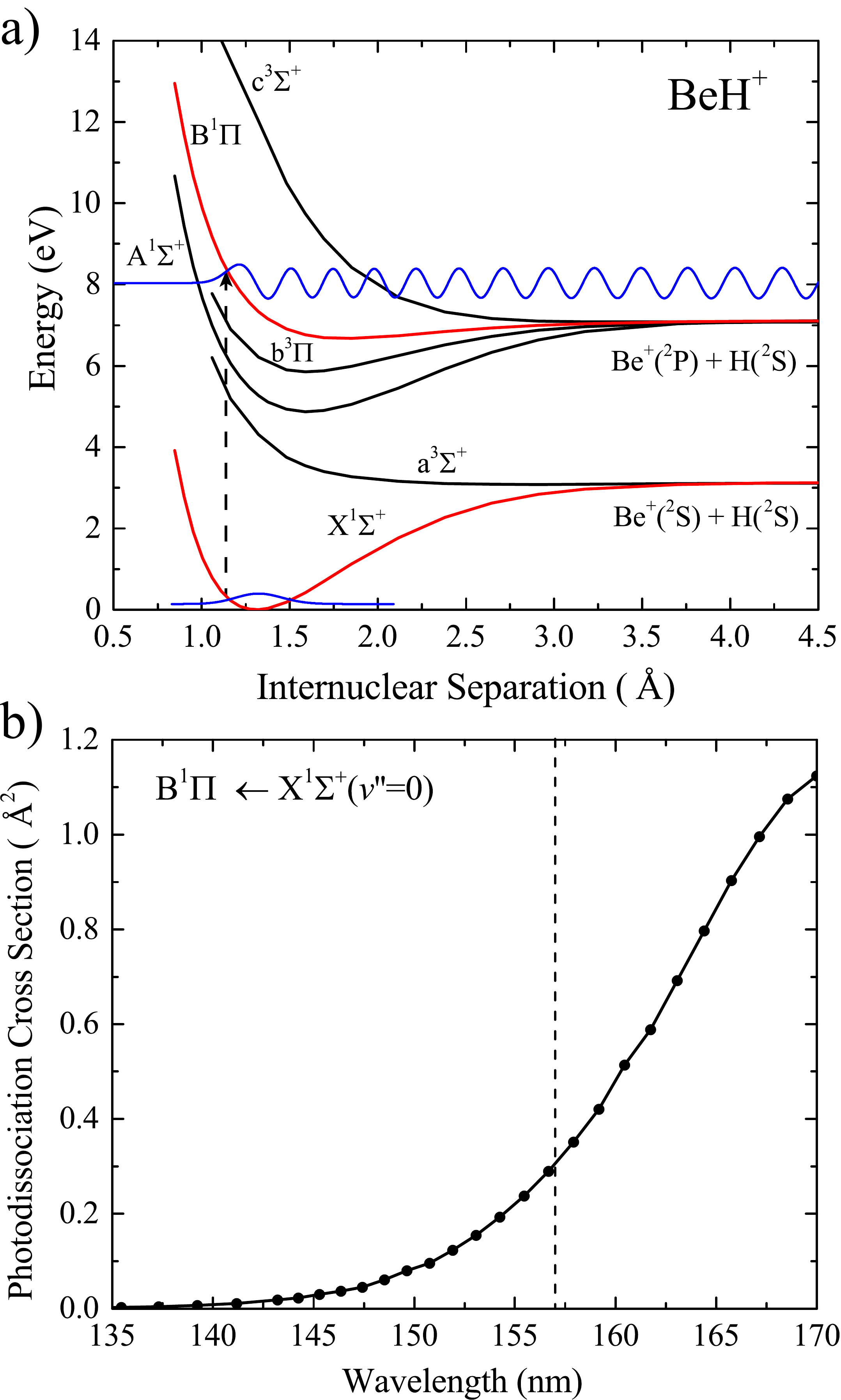}}
\caption{\label{behlevels}(color online) (a) Low-lying electronic states of the BeH$^+$ molecular ion as calculated in Refs.~\cite{Machado91,Farjallah13}. We excite the $\text{B}^1\Pi \leftarrow \text{X}^1\Sigma^+(v''=0)$ transition above the dissociation threshold to produce Be$^+$ and H as given by the $\text{B}^1\Pi$ asymptote. Relevant wavefunctions calculated as numerical solutions to the radial Schr\"{o}dinger equation are depicted in blue (not drawn to scale). The vertical dashed arrow indicates the 157 nm excitation wavelength used for this work. (b) Calculated thermally-averaged photodissociation cross section versus photon wavelength for the transition of interest. The vertical dashed line indicates the 157 nm excitation wavelength used for this work. The calculation is truncated at 170 nm, which is $\sim$0.1 eV above the dissociation threshold of the B$^1\Pi$ potential.}
\end{figure}

In Fig.~\ref{behlevels}(a), we present the six lowest-energy electronic states of the BeH$^+$ molecule as calculated in~\cite{Machado91,Farjallah13}. The given electronic states all asymptote to Be$^++$ H at large internuclear separation; the Be $+$ H$^+$ asymptotes possess larger energies. The ground state of BeH$^+$ exhibits $\text{X}^1\Sigma^+$ symmetry, which allows for electric dipole transitions to the spin-singlet excited states $\text{A}^1\Sigma^+$ and $\text{B}^1\Pi$. The minimum of the $\text{B}^1\Pi$ potential is sufficiently shifted from that of the $\text{X}^1\Sigma^+$ to provide a pathway for single-photon photodissociation from the lowest vibrational level ($v''=0$) of the ground electronic state. The BeH$^+$ ions are sympathetically cooled by the laser-cooled Be$^+$ ions. Previous studies indicate a strong sympathetic cooling of the axial motion of the BeH$^+$ with a resulting temperature of $\sim$1 mK and the formation of a centrifugally separated BeH$^+$ crystal outside of the Be$^+$ ions~\cite{Jensen05}. However the cyclotron motion of the BeH$^+$ decouples from the axial motion and is only cooled to $\sim$1 K. The BeH$^+$ internal degrees of freedom are in equilibrium with the 300 K blackbody radiation of the trap chamber. At room temperature, less than $0.01$ \% of BeH$^+$ populate excited vibrational states, but the lowest $\sim$10 rotational states are thermally populated. Resonance-enhanced multi-photon dissociation (REMPD), while useful for molecular ion spectroscopy~\cite{Roth06,Bertelsen04,Seck14,Ni14}, is rotational-state selective and would require a widely tunable laser for our goal of complete BeH$^+$ removal. In the present work, we couple the $\text{X}^1\Sigma^+$ rotational states directly to the continuum with a single photon. Vibrational wavefunctions corresponding to the relevant bound ($\text{X}^1\Sigma^+(v''=0)$) and free-particle ($\text{B}^1\Pi$) states (see Fig.~\ref{behlevels}(a)) are calculated by numerically solving the one-dimensional Schr\"{o}dinger equation with the given electronic potentials~\cite{Johnson77}.

Given the level structure of Fig.~\ref{behlevels}(a), we calculate the PD cross section for each ground rotational state, $J''$, as~\cite{Singer83}
\small
\begin{eqnarray}
\sigma(J'') &=& \frac{4 \pi^2}{e^2} \alpha \mu_{trans}^2 \nonumber \\
&\times& E_{ph} \left| \int_{0}^{\infty} \phi(E,R)\psi_{v''=0}(R) dR \right|^2 \nonumber \\
&\times& \sum_{J'm_J'm_J''} \frac{\left| \langle \Pi,J',m_J' |\hat{T}^{(1)}_{-1}+\hat{T}^{(1)}_{+1}|\Sigma,J'',m_J''\rangle \right|^2}{2(2J'' + 1)}, \label{sigma}
\end{eqnarray}
\normalsize
where $\alpha$ is the fine structure constant, $e$ ($a_0$) is the electron charge (Bohr radius), $\mu_{trans}=1.136 \: ea_0$ is the $\text{B}^1\Pi \leftarrow \text{X}^1\Sigma^+$ transition dipole moment calculated in~\cite{Machado91,Farjallah13}, $E_{ph}$ is the photon energy, $\phi(E,R)$ is upper-state vibrational wavefunction with energy $E$ above the dissociation threshold, $\psi_{v''=0}(R)$ is the bound state vibrational wavefunction, and $R$ is the internuclear separation. A double-prime indicates a lower state and a single-prime indicates an upper state throughout this report. The $\hat{T}^{(1)}_{\pm 1}$ operators are dimensionless rank-1 spherical tensors signifying electric dipole coupling with linearly-polarized photons propagating along $z$ (see Fig.~\ref{trap}(a)). We present Eq.~(\ref{sigma}) such that the first line describes the electronic degree of freedom; the second and third lines are both dimensionless and represent vibrational and rotational degrees of freedom, respectively.

To calculate the vibrational overlap of the bound and free states (i.e. Franck-Condon density) over a range of excitation wavelengths, we numerically solve the one-dimensional Schr\"{o}dinger equation for each upper-state energy, $E$, and subsequently compute the overlap of each with the ground vibrational wavefunction. Finally, to produce the cross section curve of Fig.~\ref{behlevels}(b) we perform a thermal average over rotational states $0 \leq J'' \leq 20$ as
\begin{equation}
\langle \sigma (J'') \rangle_{th} \equiv Z^{-1}\sum_{J''=0}^{20}(2J''+1)\sigma(J'')\exp{\left(-E_{J''}/k_bT\right)}. \label{therm}
\end{equation}
In Eq.~(\ref{therm}), $Z$ is the partition function, $k_b$ is Boltzmann's constant, $T$ is room temperature, and $E_{J''}$ is the relative energy of level $J''$. Note that the $\text{B}^1 \Pi$ state is predicted to support at least eight bound vibrational levels~\cite{Machado91,Farjallah13}, but we neglect any off-resonant couplings to such bound states for this work as the long-wavelength limit of the cross section calculation of Fig.~\ref{behlevels}(b) is $\sim$0.1 eV above the dissociation energy of the B$^1\Pi$ potential. We additionally neglect the Penning trap magnetic field and BeH$^+$ hyperfine structure in calculating state energies; both are small compared to the breadth of the photodissociation cross section of Fig.~\ref{behlevels}(b).

Figure~\ref{behlevels}(b) suggests a photon wavelength of $\sim$170 nm for maximum PD rate. However, given the technical difficulty of making laser sources in the vacuum ultraviolet (VUV), we use a commercial excimer laser operating with F$_2$ gas to produce 157 nm pulses with 5 ns width and a repetition rate of 500 Hz at $\leq$2 mJ per pulse. Molecular oxygen absorbs strongly near 160 nm, thus the entire laser beam path is purged with high-purity N$_2$ gas so that only $\sim$10 \% of the beam energy is absorbed before entering the trap vacuum system through a CaF$_2$ viewport. We measure an exponential decay length of $\sim$8.2 m with the N$_2$ purge gas, while the out-of-vacuum path length from excimer laser to CaF$_2$ viewport is 1.25 m. As depicted in Fig.~\ref{trap}, the excimer beam is aligned to the parallel Doppler cooling laser beam and propagates in the opposite direction along the trap magnetic field to the ion crystal. We have installed a normal-incidence mirror with a 157 nm reflectance of $\sim$88 \% below the trap electrodes in order to shield a fused silica viewport at the bottom of the trap vacuum apparatus from the high-intensity VUV illumination. This dielectric-coated mirror transmits $\sim$89 \% of the 313 nm cooling beam intensity. Ideally the excimer beam should be aligned such that it is reflected back along the incoming beam path, exiting the vacuum chamber through the CaF$_2$ viewport. Geometric constraints with our current trap chamber preclude this; these constraints also cause partial blockage of the excimer beam before it enters the experimental trap zone. In the absence of blockage, we calculate a rectangular excimer beam waist at the ions of $\sim$1.7 mm $\times$ 0.85 mm due to our use of a CaF$_2$ spherical lens with a focal length of 85 cm placed a distance of 1.3 m upstream from the ion crystal. This beam waist combined with the 2 mJ pulse energy gives an average intensity of 21 W cm$^{-2}$.

Figures~\ref{trap}(b) and~\ref{trap}(c) show images of Be$^+$ ion crystals with $\omega_r = 2\pi \times 110$ kHz (b) before and (c) after a photodissociation sequence of 9000 pulses at 1.6 mJ each. Three-dimensional ion crystals form elliptical boundaries in the Penning trap potential of Eq.~(\ref{trappot}). The outer edges of the crystal of Fig.~\ref{trap}(b) are dark due to heavier-mass impurity ions that centrifugally separate under the rotation of the crystal. After photodissociation, we observe that the elliptical boundary is restored with fluorescing Be$^+$ and the total volume of the crystal is unchanged, indicating a conservation of total ion number. Fits to the Be$^+$ fraction in the images of Figs.~\ref{trap}(b) and~\ref{trap}(c) show that we have increased from 81 \% to 97 \% Be$^+$, and cyclotron resonance data reveals the presence of BeO$^+$ and BeOH$^+$ molecules in the remaining dark regions of Fig.~\ref{trap}(c)~\cite{PDfoot}. We note that no protons are observed within the Coulomb crystal after PD, thereby confirming the asymptote of the theoretical B$^1\Pi$ potential of Fig.~\ref{behlevels}(a)~\cite{Machado91,Farjallah13}.

\begin{figure}[t]
\resizebox{8.0cm}{!}{
\includegraphics{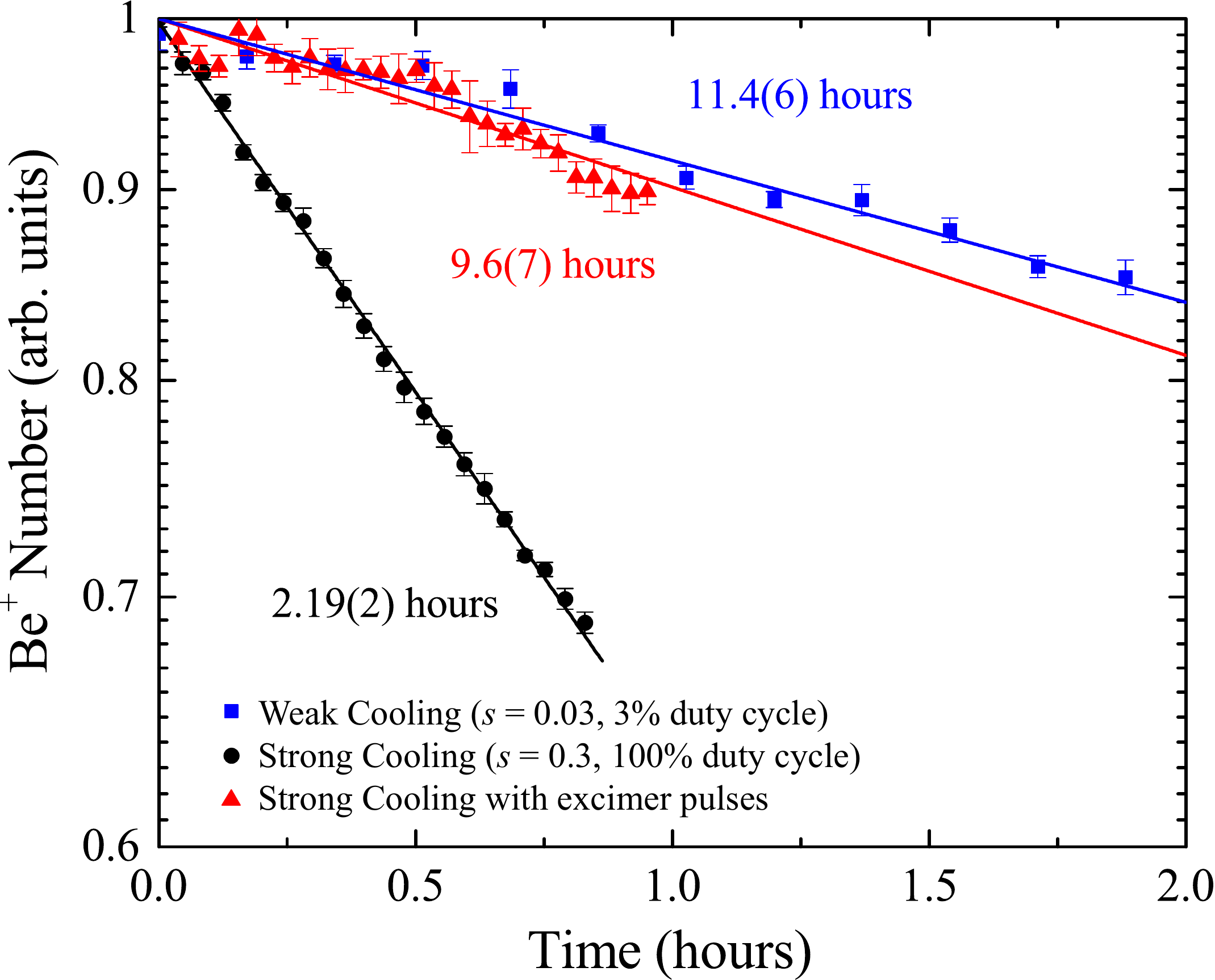}}
\caption{\label{pdsummary}(color online) Measured (points with error bars) and fit (solid lines) Be$^+$ exponential decay plotted on a logarithmic vertical scale. Each of the three data sets consists of one experimental run, and vertical error bars represent one standard deviation. With a saturation parameter, $s$, of 0.03 and a 3 \% duty cycle (`weak cooling'), we measure a Be$^+$ lifetime of 11.4(6) hours that is limited by chemistry with background gas molecules other than H$_2$ ({\color{blue}$\blacksquare$}). Using a $>300\times$ larger photon flux for Doppler cooling ($\bullet$), the Be$^+$ number decays with a 2.19(2)-hour time constant that is consistent with an H$_2$ background pressure of $7 \times 10^{-9}$ Pa ($5 \times 10^{-11}$ Torr). To mitigate BeH$^+$ production, we trigger 200 pulses (500 Hz repetition rate) of a 157 nm excimer laser every 2 min. ({\color{red}$\blacktriangle$}). The resulting rate of BeH$^+$ photodissociation is sufficient to increase the Be$^+$ lifetime by a factor of 4.4(3) to be roughly consistent with the `weak cooling' measurements.}
\end{figure}

To demonstrate the efficacy of this BeH$^+$ photodissociation technique for quantum information experiments, we measure single-exponential Be$^+$ number decay due to conversion to impurity ions under three different conditions as shown in Fig.~\ref{pdsummary}: `weak' Doppler laser cooling ({\color{blue}$\blacksquare$}), `strong cooling' ($\bullet$), and `strong cooling' in the presence of photodissociation pulses ({\color{red}$\blacktriangle$}). In all cases, relative Be$^+$ number corresponds to the Doppler cooling fluorescence count rate as measured on an ultraviolet-sensitive photomultiplier tube. We measure a maximum Be$^+$ lifetime of 11.4(6) h by minimizing the parallel Doppler cooling photon flux to a 3 \% duty cycle and $s=0.03$ saturation parameter, thereby producing BeH$^+$ impurities with a predicted time constant $>$200 h. The significantly shorter 11.4(6) h measured lifetime appears to be due to Be$^+$ collisions with background gas molecules that produce heavier impurity ions in the absence of any Doppler cooling light. Neutral alcohol molecules such as acetone ((CH$_2$)$_3$CO) and/or methanol (CH$_3$OH), both of which were used to clean the vacuum components, undergo exothermic reactions with ground-state Be$^+$, and Langevin cross section estimates presented in the Appendix suggest that partial pressures of $\sim10^{-11}$ Pa ($10^{-13}$ Torr) could explain both the observed conversion rate and the detected BeO$^+$ and BeOH$^+$ impurities~\cite{PDfoot}.

After increasing the Doppler cooling parameters to $s=0.3$ at 100 \% duty cycle, the Be$^+$ trap lifetime drops to 2.19(2) h ($\bullet$) -- consistent with a background H$_2$ pressure of $\sim$ $7 \times 10^{-9}$ Pa ($\sim$ $5\times10^{-11}$ Torr) as BeH$^+$ production dominates all other Be$^+$ conversion rates. The unit duty cycle for Doppler cooling is representative of typical trapped-ion quantum information work where the time for cooling and preparing initial qubit states is a significant fraction of the experimental sequence~\cite{Home09,Britton12,Kim11}. To demonstrate PD, we apply 200-pulse bursts of the excimer laser every 2 minutes with a pulse energy of 1.9 mJ and repetition rate of 500 Hz in the regime of `strong cooling' ({\color{red}$\blacktriangle$}). The total time for each pulse burst is 0.4 s, and the Doppler cooling is switched off for this duration as well as 2 s after the photodissociation sequence, yielding a negligibly-modified strong cooling duty cycle of $\sim$98 \%. The PD pulses yield an increased time constant for decay of 9.6(7) h, which is roughly consistent with the `weak cooling' result. The small difference between the 11.4 h and 9.6 h time constants is likely the result of daily fluctuations in background gas pressure. A high rate (i.e. every 2 min.) of PD pulses is necessary to maintain a pure Be$^+$ crystal since we observe that BeH$^+$ converts to BeO$^+$ and BeOH$^+$ at a rate three times faster than Be$^+$($^2$P) + H$_2 \rightarrow$ BeH$^+ +$ H conversion under `strong cooling' conditions.

In conclusion, we have demonstrated the first photodissociation results exploiting the $\text{B}^1\Pi \leftarrow \text{X}^1\Sigma^+$ electronic transition in BeH$^+$. This technique removes an important limitation for large-scale quantum information experiments that rely on Be$^+$ and does so with a commercially-available, turnkey laser system. We achieve a more than four-fold increase in experimental run-time with a 2 \% reduction of the duty cycle. Improved beam alignment would enable an experimental determination of the absolute PD cross section at 157 nm as well as optimization of the dissociation beam profile. We predict a maximum BeH$^+$ PD rate of $\sim$1.2 pulse$^{-1}$ from our estimated PD beam average intensity of 21 W cm$^{-2}$ and the theory results of Fig.~\ref{behlevels}(b). The observed PD rate is lower presumably due to partial excimer beam obstruction along the 1.3-m in-vacuum path. We estimate that a 193 nm ArF excimer laser will dissociate MgH$^+$~\cite{Aymar09} and AlH$^+$~\cite{Seck14} with dissociation cross sections of $\sim$0.4 \AA$^2$ and $\sim$$10^{-3}$ \AA$^2$, respectively, making this technique applicable for removing a number of experimentally-relevant impurity species.

This work was supported by NIST. J. G. Bohnet is supported by an NRC fellowship funded by NIST. The authors thank E. R. Hudson, K. Chen, and J. L. Bohn for useful discussions as well as D. T. C. Allcock and Y. Wan for comments on the manuscript. This manuscript is a contribution of NIST and not subject to U.S. copyright.

\section*{Appendix} \label{appendix}
\subsection*{Estimating $C_4$ Coefficients for Polar Molecule-Ion Chemistry}

As described in~\cite{Bible,BellSoftley}, we define the Langevin `capture' cross section between an ion and polarizable neutral particle as
\begin{equation}
\sigma_L = 2\pi \sqrt{\frac{C_4}{E_c}}, \label{sigma4}
\end{equation}
where $E_c$ is the ion-neutral collision energy and $C_4$ is the coefficient of the attractive interaction potential between scatterers whose magnitude is proportional to $R^{-4}$, where $R$ is the particle separation. Equation~\ref{sigma4} reveals the unique property of $C_4$ collisions that the Langevin rate,
\begin{equation}
k_L = \sigma_L v_{rel} = 2\pi \sqrt{\frac{2C_4}{M}},
\end{equation}
is independent of collision energy, where $v_{rel}(M)$ is the relative velocity (reduced mass) of the scatterers. To obtain the $C_4$ coefficient for ion-neutral collisions, we begin with the energy shift ($W$) of a particle with polarizability $\alpha_{pol}$ in the electric field ($E$) of an ion:
\begin{eqnarray}
W &=& \frac{1}{2} \alpha_{pol} |E|^2 \\
&=& \frac{\alpha_{pol}}{2} \frac{e^2}{(4\pi\epsilon_0)^2 R^4}. \label{Wsi}
\end{eqnarray}
From Eq.~\ref{Wsi} we obtain the $C_4$ coefficient necessary for evaluating $k_L$:
\begin{equation}
C_4 = \frac{\alpha_{pol} e^2}{2 (4\pi\epsilon_0)^2}.
\end{equation}

For colliding molecules with a permanent electric dipole moment ($\mu$) as defined in the molecule-fixed frame, one can calculate the polarizability ($\alpha_{JK}$) from the second-order Stark shift for a given rotational level ($J$) and projection ($K$) given in~\cite{TownesSchawlow}:
\begin{widetext}
\begin{equation}
\alpha_{JK} = \frac{1}{2J+1} \frac{\mu^2}{B} \sum_{m_J=-J}^J  \left[ \frac{(J^2 - K^2)(J^2-m_J^2)}{J^3(2J-1)(2J+1)} - \frac{[(J+1)^2 - K^2][(J+1)^2 - m_J^2]}{(J+1)^3 (2J+1) (2J+3)}\right] \label{alphaJK}.
\end{equation}
\end{widetext}
In the above equation, each $J$ level is coupled to its two nearest neighbors and $B$ is the rotational constant. The quantum number $K$ ($m_J$) is the projection of $J$ along the molecule symmetry (external field) axis. Both $K$ and $m_J$ may take the values $\{ -J, -J+1,...,J-1,J \}$, and Stark shifts for opposite-sign projections are identical under an external electric field. Equation~\ref{alphaJK} represents the polarizability of the level $(J,K)$ averaged over all possible projections, $m_J$.

Specializing to methanol (CH$_3$OH, $\mu=0.67 \: ea_0$), we begin with the rotational constants $A = 4.222$ cm$^{-1}$, $B = 0.822$ cm$^{-1}$, and $C = 0.792$ cm$^{-1}$. Since $B\sim C$, methanol is a slightly asymmetric prolate top. For this analysis, we treat methanol as a pure symmetric top whose energy levels, $U_{JK}$, are given by:
\begin{equation}
U_{JK} = \left( \frac{B+C}{2} \right) J(J+1) + \left(A - \frac{B+C}{2} \right)K^2.
\end{equation}
Since background molecules in the vacuum chamber are assumed to be at room temperature, we must compute a temperature-averaged polarizability if many rotational states lie below $k_b T \sim 209$ cm$^{-1}$. With the usual Boltzmann weighting, we obtain
\begin{equation}
\langle \alpha_{JK} \rangle_{th} = Z^{-1} \sum_{J=0}^{50} (2J+1) \sum_{K=-J}^{+J} e^{-\frac{U_{JK}}{k_b T}}\alpha_{JK}
\end{equation}
where $Z$ is the partition function. Including all states up to $J=50$, we obtain the following average polarizability for methanol:
\begin{equation}
\langle \alpha_{JK} \rangle_{th} \sim 2.23 \times 10^{-39} \text{ C m$^2$ V$^{-1}$}.
\end{equation}
Using the above thermally-averaged polarizability, we compute the Langevin reaction rate for Be$^+$-methanol collisions:
\begin{eqnarray}
k_L &=& 2\pi \sqrt{\frac{e^2 \langle \alpha_{JK} \rangle_{th}}{M (4 \pi \epsilon_0)^2}} \\
&\sim& 4 \times 10^{-9}  \text{ cm$^3$ s$^{-1}$}.
\end{eqnarray}
With a Be$^+$ decay time constant of $\tau = 11.4$ hr., we extract a background methanol density estimate of:
\begin{eqnarray}
\rho_{methanol} &=& (\sigma_L v_{methanol} \tau)^{-1} \\
&\sim& 6.2 \times 10^{3} \text{ cm}^{-3} \\
&\sim& 2\times 10^{-13} \text{ Torr}.
\end{eqnarray}


\end{document}